\documentclass[iop]{emulateapj}

\usepackage{epstopdf}
\usepackage{graphicx}
\usepackage{natbib}
\usepackage{color}

\shorttitle{STRUCTURES IN THE MAGNETOSHEATH}
\shortauthors{V\"or\"os et~al.}

\begin{document}


\title{Turbulence generated proton-scale structures in the terrestrial magnetosheath}
\author{Zolt\'an V\"or\"os\altaffilmark{1,2} \and Emiliya Yordanova\altaffilmark{3} \and Marius M. Echim\altaffilmark{4,5} \and Giuseppe Consolini\altaffilmark{6} \and Yasuhito Narita\altaffilmark{1}}
\altaffiltext{1}{Space Research Institute, Austrian Academy of Sciences, Graz, Austria} \email{zoltan.voeroes@oeaw.ac.at}
\altaffiltext{2}{Department of Geophysics and Space Sciences, E\"otv\"os University, Budapest, Hungary}
\altaffiltext{3}{Swedish Institute of Space Physics,Uppsala, Sweden}
\altaffiltext{4}{Belgian Institute for Space Aeronomy, Bruxelles, Belgium}
\altaffiltext{5}{Institute for Space Sciences, M\^agurele, Romania}
\altaffiltext{6}{INAF-Istituto di Astrofisica e Planetologia Spaziali, Roma, Italy}





\begin{abstract}
Recent results of numerical magnetohydrodynamic simulations suggest that in collisionless space plasmas turbulence can spontaneously generate thin current sheets. These coherent structures can partially explain intermittency and the non-homogenous distribution of localized plasma heating in turbulence. In this Letter Cluster multi-point observations are used to investigate the distribution of magnetic field discontinuities and the associated small-scale current sheets in the terrestrial magnetosheath downstream of a quasi-parallel bow shock. It is shown experimentally, for the first time, that the strongest turbulence generated current sheets occupy the long tails of probability distribution functions (PDFs) associated with extremal values of magnetic field partial derivatives. During the analyzed one hour long time interval, about a hundred strong discontinuities, possibly proton-scale current sheets were observed.
\end{abstract}





\section{Introduction}
Downstream of the terrestrial bow shock (BS) the super-sonic and super-Alfv\'enic solar wind flow slows down, gets compressed and heated.
The solar wind is diverted by the strong geomagnetic field at the magnetopause (MP). The magnetosheath (MS) is the region between the BS and the MP, where the field and plasma fluctuations are rather strong. Numerical simulations \citep{Omidi14, Karimabadi14} demonstrate the complexity of this region - being populated by various interacting multi-scale structures, such as filaments, vortices, current sheets, plasma jets and flows. In-situ measurements from different missions also revealed the existence of nonlinear structures in upstream and downstream BS regions, for example,  shocklets \citep{Hoppe81}, short duration large amplitude magnetic structures (SLAMS) \citep{Lucek04}, and hot-flow anomalies \citep{Facsko09} embedded in the highly turbulent MS. Some of these nonlinear structures are also associated with enhanced levels of wave and fluctuation activity \citep{Kovacs14}.
The fluctuations and structures are more pronounced in the quasi-parallel BS configuration when the angle between the interplanetary magnetic field (IMF) and the nominal BS normal is smaller than  45$^\circ$.
In such a case, the upstream (solar wind, foreshock) and downstream MS regions are magnetically connected and the resulting turbulence becomes increasingly intermittent away from the BS \citep{Yordanova08}.

In turbulent space plasmas thin magnetic structures, current sheets and reconnection can be generated spontaneously through complex interactions
\citep{Chang04, Servidio09, Matthaeus15}. Furthermore, recent studies on the statistics of velocity gradient tensor invariants
revealed that approaching the non-MHD scales vortex stretching may play a relevant role in generating small scale dissipative structures \citep{Consolini15}. These structures observed as discontinuities in the solar wind \citep{Greco09} are associated with local dissipation/heating \citep{Osman12, Osman14} and introduce intermittency to plasma turbulence.
The occurrence frequency of discontinuities in the solar wind agrees rather well with the statistics of turbulence generated current sheets in numerical simulations \citep{Greco08, Greco09, Servidio11}. Cluster observations in MS downstream quasi-parallel BS also confirmed the occurrence of reconnecting thin current sheets \citep{Retino07,Sundkvist07} and the current sheet associated electron heating \citep{Chasapis15a}.

Numerical simulations have shown that the strongest discontinuities and current sheets generated by turbulence populate the tails of non-Gaussian PDFs of normalized current density \citep{Greco09, Matthaeus15}. However, this has not been shown directly from the data. Although it is tempting to assume that the fat tails of the PDFs are completely determined by turbulence generated strong current sheets, one should not forget that there might exist other intermittent structures of different origin. As a matter of fact, the high Reynolds number simulations of MHD turbulence \citep{Greco09, Servidio11} cannot reproduce the rich ensemble of intermittent nonlinear structures mentioned above. In this letter, using Cluster data in the MS downstream of quasi-parallel BS, we show that the strongest observed current sheets de facto occupy the tails of histograms. We also estimate the occurrence frequency of the strongest current sheets along the trajectory of Cluster spacecraft.
The paper is organized as follows: Section 2 explains the data and instrumentation, Section 3 introduces the structure detection tools. Section 4 demonstrates how the structure detection tools work for a thin current sheet. In Section 5 discontinuity and current sheet statistics is presented. Section 6 contains the summary and conclusions.

\section{Data and instrumentation}

In this Letter we consider the data interval between $9.6$ and $10.6$ UT (decimal hours) on 27 March 2002 when the Cluster spacecraft probed the magnetosheath downstream of a quasi-parallel bow shock. OMNI $1-$minute data have been used to obtain the model bow shock \citep{Farris94} and magnetopause \citep{Shue98} boundaries as well as the normal vector to the bow shock.
The Cluster fleet was initially at the GSE coordinate system position ($10,-8.2,-8$) $R{_E}$, moving $\sim 1R{_E}$ towards the magnetopause in a nearly perfect tetrahedron configuration with inter-probe separations of about 100 km. The relative positions of the spacecraft are shown in Figure $1$. The upstream plasma and field conditions in the solar wind were quasi-stationary with average parameters - magnetic field $\sim 3.4$ nT, velocity $\sim 446$ km/s, and density $\sim 3$ cm$^{-1}$.

During the selected interval, the magnetic field measurements were available from the FGM instrument \citep{Balogh97} with sampling frequency $f_s=67$ Hz, and from the STAFF instrument \citep{Cornilleau-Wehrlin97} with $f_s=450$ Hz. By using finite impulse response filter at $\sim1$ Hz the magnetic data from both instruments were merged into one time-series with extended frequency coverage up to 450 Hz in burst mode. The electric field data were available from EFW instrument \citep{Gustafsson97}.

\section{Structure detection tools}
Current sheets, associated with large-scale boundaries, such as the magnetopause or the magnetotail current sheet can be easily identified. Even if the boundaries are moving or the structures are flapping, the same current sheet can be observed multiple times. On the other hand, turbulence generated boundaries and the associated current sheets are small-scale and transient in the spacecraft frame. Their identification requires special methods.
To detect discontinuities and current sheets in magnetosheath three types of parameters will be estimated from Cluster spacecraft pairs ($Ci-Cj$, $i,j$= 1, 2, 3, 4) in time $t$:  1.) the Partial Variance of Increments ($PVI_{ij}(t)$); 2.) the angle between magnetic field vectors $\theta_{ij}(t)$; and 3.) the magnetic field derivatives $\partial_{ij}(t)$.

The method using PVIs was first introduced for identification of discontinuities within the intermittent solar wind turbulence from single point measurement. The results have also been supported by numerical simulations \citep{Greco09}.
The PVIs here are calculated on the basis of the normalized variance of the absolute value of magnetic field spatial increments between two spacecraft
$\Delta \textbf{\textit{B}}_{ij}(t)= \textbf{\textit{B}}_i(t) - \textbf{\textit{B}}_j(t)$: 

\begin{equation}
PVI_{ij}(t)=\sqrt{\frac{\mid{\Delta \textbf{\textit{B}}_{ij}(t)\mid}^2}{\big \langle{\mid{\Delta \textbf{\textit{B}}_{ij}}\mid ^2}\big \rangle}},
\label{eq1}
\end{equation}
where the average $\langle \cdot \rangle$ is taken over the whole interval. The mean $PVIm(t) =\sum_{ij}PVI_{ij}(t)/6$ is also calculated. In contrary to the standard one-directional $PVI$ \citep{Greco08}, $PVIm$ comprises information on increments along all $Ci-Cj$ pairs in Cluster tetrahedron. In recent 3D simulations of MHD turbulence a similar multi-directional PVI was used, arguably providing an unbiased information about the spatial structure of discontinuities \citep{Zhang15}.

The rotation of the magnetic field vector between spacecraft pairs is estimated through:
\begin{equation}
\theta_{ij}(t)=\cos^{-1}\frac{\textbf{\textit{B}}_i(t)\cdot \textbf{\textit{B}}_j(t)}{\mid\textbf{\textit{B}}_i(t)\mid \cdot \mid\textbf{\textit{B}}_j(t)\mid}.
\label{eq2}
\end{equation}

The partial derivatives of the magnetic field are obtained from:
\begin{equation}
\partial_{ij}(t)\equiv\frac{\Delta \textbf{\textit{B}}_{ij}(t)} { \Delta \textbf{\textit{r}}_{ij}(t)}
\label{eq3}
\end{equation}
where $\Delta \textbf{\textit{r}}_{ij}(t)$ is the spatial separation between $Ci,Cj$ pairs, $\partial_{ij}$ stands for the partial derivatives $\partial B_X/\partial X$, $\partial B_X/\partial Y$, $\partial B_X/\partial Z$, etc.
These form the orthogonal components of $(\nabla \times \textbf{\textit{B}})_{xyz}$ and $(\nabla \cdot \textbf{\textit{B}})_{xyz}$.
For supposedly time-independent short events with roughly constant
$(\nabla  \times \textbf{\textit{B}})_{xyz} $
and for linearly changing magnitudes of magnetic components over the tetrahedron, the current density
$\textbf{\textit{j}}=(\nabla \times \textbf{\textit{B}}) / \mu_{0} $ ($\mu_{0}$ is the magnetic constant) can be calculated.


It has already been shown that the parameters $PVI$ and $\theta$ are correlated. Discontinuities with high values of $PVI$ are associated with large magnetic rotations $\theta > 90^{\circ}$ \citep{Chasapis15a}. The joint occurrence of strong magnetic shear and high $PVI$ indicates that the corresponding discontinuity is a current sheet. More about structure detection tools in turbulence can be found in \citet{Chasapis15b}.
For the studied interval the correlations are demonstrated in Figure $2$.
The left panels $2a.$--$2d.$ show the magnetic field intensity for C1, the PVI magnitude for the spacecraft pair C1--C4, the magnetic field rotation angle for the same pair of probes and the partial derivatives for the $B_Z$
components of the magnetic field. The parameters from other spacecrafts or spacecraft pairs look similar (not shown). The right panels $2e.$--$2h.$ show the PDFs corresponding to the time series on the left.
The maxima of PDFs are normalized to one. Gaussians with the same standard deviations are inserted as well. Although the PDF of B(C1) is the closest to the Gaussian distribution, there are significant deviations from
it near the peak value. It indicates, that various physical processes with different magnetic PDFs can form the summary histogram within the analyzed interval. 
The large deviations from the Gaussians for the other parameters are evident near the maxima and at the tails of PDFs.
Since the strongest discontinuities or (reconnecting) current sheets are expected to form the tail of PDFs \citep{Greco09}, we are interested in correlations between the extremal values of $PVI_{ij}$, $\theta_{ij}$ and $\partial_{ij}$. The green boxes in panels $2a.$--$2d.$ show a few sub-intervals of the data when the threshold $PVI(C1-C4)=4$ selects intervals of large-deviations of $\theta_{ij}$ and $\partial_{ij}$ as well.

\section{An example of current sheet}
Figure $3$ shows a crossing of thin proton-scale ($\sim$0.5 sec) current sheet by the Cluster fleet, for which \citet{Chasapis15a} found electron heating signatures. For this event the GSE coordinate system was shown to be very close to the current sheet system. The panels show the bipolar change of the electric field $E_X(C2)$ ($3a.$), the $B_Z (C1, C2, C3, C4)$ components of the magnetic field ($3b.$), $PVI$s for spacecraft pairs and the mean $PVIm$ ($3c.$), $\theta$s for spacecraft pairs ($3d.$) and the derivatives $\partial_{ij}$ ($3e.$--$3g.$). The bipolar $E_X(C2)$ signature is associated with the fast, $\sim$0.06 sec long sign-change of $B_Z(C2)$ within the dashed box. The average Doppler-shifted electron gyroperiod for the considered 1 h long interval is $\sim$ 0.04 sec, rather close to the duration of the bipolar electric field change. This thin electron current sheet is embedded into a broader sheet associated with the more gradual changes of $B_Z$ before and after of the dashed box. The fastest sign-changes of $B_Z$ in Figure 3b were observed by $C1$ and $C2$ indicating that these probes crossed the current sheet where it was locally thinner or in a direction more perpendicular than the other probes. Since the $C1-C2$ probes are separated mainly in Y and Z directions (Figure 1), the current sheet normal direction is close to the GSE X direction. This is also confirmed by minimum variance analysis (not shown). Therefore, the normal component of the electric field is $E_n \sim E_X$ (Figure $3a.$) exhibiting the same time scale as the embedded electron scale seen in $B_Z$ (red curve in the dashed box, Figure $3b.$).
Figures $3c.$ and $d.$ show that the $PVI_{ij}$ or $PVIm$  and $\theta_{ij}$ parameters are correlated and indicate the crossings of the thin current sheet rather well. $\theta_{ij} \sim 180^{\circ}$ implies that the current sheet is between a spacecraft pair while $\theta_{ij} \sim 0^{\circ}$ indicates that the probes are at the same side of the sheet.
The derivatives of the $B_Z$ component mark the presence of the current sheet (Figure $3g.$), the other $\partial_{ij}$s show only insignificant fluctuations (Figures $3e., f.$). Some events during the analyzed interval are associated with the sign-changes and enhanced gradients of the $B_Y$ component.

\section{Current sheet and discontinuity statistics}
The potential discontinuities in the data can be found by using the mean $PVIm$.
In any case, however, the PVI thresholds should not be considered as unique parameters identifying the thin current sheets.
For the selected cases here we considered all the parameters shown in Figure 3 and visually checked the events.
The usefulness of $PVIm$ is demonstrated in Figure $4g.$, where three different thresholds (red, blue and black horizontal lines)
and the corresponding time instants - points - are shown. The red, blue and black populations of points correspond to the $PVIm$ thresholds of the same color.
The first five discontinuities are numbered and color coded by red, magenta, blue, green and brown. This color code is also used in Figure 5 for indicating the location of discontinuities in a histogram. The 1$^{st}$ discontinuity (magnetic components are depicted in Figures $4d.$--$f.$) is a magnetic reconnection event described thoroughly by \citep{Retino07}. The 2$^{nd}$ discontinuity is a current sheet with $B_Y$ sign-changes and the 3$^{rd}$ discontinuity is a more complicated event comprising crossings of two neighboring current sheets. The 4$^{st}$ discontinuity (Figures $4a.$--$c.$) was identified as a current sheet associated with electron heating by \citet{Chasapis15a}. For these current sheets the maxima of $\theta_{ij}$ are close to  $180^{\circ}$. The 5$^{th}$ discontinuity is a current sheet where the maxima of
$\theta_{ij}$ are less than  $120^{\circ}$, which is still a very high magnetic shear angle \citep{Chasapis15a}.

Figure $4h.$ shows the largest thickness of discontinuities in seconds for changing thresholds $PVIm$. It is calculated as a local duration of a discontinuity in time having $PVIm \geq threshold$. Since for a given threshold a discontinuity may be represented by one \textbf{$PVIm \geq threshold$} value only, the smallest thicknesses are often close to 0 second. Therefore, the mean discontinuity thickness has no meaning. However, the largest thickness (Figure $4h.$) together with the total number of detected discontinuities (Figure $4i.$) for a given $PVIm$ threshold provides the information about the minimum number of discontinuities with thicknesses equal or smaller than the largest thickness. The comparison of Figure $4h.$ with Figure $4i.$ shows, that for very thin structures, for example with a duration of less than one second (possibly current sheets), there exist almost one hundred discontinuities during the analyzed magnetosheath interval. Nevertheless, Figure $4g.$ demonstrates that the strongest discontinuities represent a tiny part of the time series only. In order to show that the intermittently occurring rare current sheets belong to the tails of PDFs as it is expected from 2D MHD simulations of current density distributions \citep{Greco09} the PDF($\textbf{\textit{j}}$) should be obtained directly from the data.
Although non-Gaussian skewed distributions are typical for space plasmas \citep{Burlaga98, Voros15}, the direct estimation of $\textbf{\textit{j}}$ from in-situ data is difficult. The curlometer technique provides an estimate of $(\nabla \times \textbf{\textit{B}})$, but it is loaded by several known sources of errors \citep{Vallat05}. The curlometer works satisfactorily for well selected, short and stationary events with a linear variation of $B$ inside the tetrahedron. This is not the case for the entire one-hour long interval analyzed here.
To reconstruct the whole PDF($\textbf{\textit{j}}$), the localized strong currents (PDF tail), the nearly current-free flux tube regions and the random transient currents (central part of PDF) should all be involved in a histogram.
However, the quality of the current density estimation for the considered time interval is low. This can be deduced from the ratio $div\textbf{\textit{B}} / |\nabla \times \textbf{\textit{B}}|$, which has values $\gg 1$ for large parts of the data. For good quality ($\textbf{\textit{j}}$) estimations it  should be $\ll 1$ \citep{Grimald12}.
This is why instead of the derived quantity ($\nabla \times \textbf{\textit{B}}$), which can also be burdened with additional errors, we use the partial derivatives $\partial_{ij}$ for further statistical investigations. However, as Figures $3e.$--$g.$ show, some of the $\partial_{ij}$s are close to zero even during current sheet crossing. On the other hand, there exist nonzero $\partial_{ij}$ values which are not associated with the current sheet. The simplest way to obtain PDF($\partial_{ij}$) is to take the extreme value of $\partial_{ij}(extr,t) = max(\partial_{ij}(t))$ or $= min(\partial_{ij}(t))$ at times $t$.
Physically, we select structures which are associated with the largest magnetic gradient between spacecraft pairs in the tetrahedron volume at times $t$.
The result is shown in Figure $5$. The black points correspond to the histogram of the entire time series of $\partial_{ij}(extr,t)$. The left (negative values) and the right (positive values) of PDF($\partial_{ij}(extr)$) are obtained independently. Since the extreme value distribution has very low occurrence frequencies for small values of derivatives, the central noisy part of the histogram for $-0.05 < \partial_{ij}(extr) < 0.05$ [nT/km] is cut off. The color coded points on the PDF curve, according to their $\partial_{ij}(extr)$, correspond to the five current sheets for different thresholds $PVIm$ shown in Figure $4g.$.
We note, that the boxes in histograms near a given value of $\partial_{ij}(extr)$ contain contributions from many other time intervals. Nevertheless, for the purposes of this study, it was enough to show that the strong discontinuities belong to the histogram boxes at the tails of PDFs.
The green curves in Figure $5$ correspond to Gaussian distributions. The smallest values are normally distributed, while at the non-Gaussian tail the strongest discontinuities are observed as it is expected from numerical simulations \citep{Greco09, Servidio09, Matthaeus15}. The wider Gaussian in Figure $5$ has the same mean and standard deviation as the data. The narrower Gaussian represents a fit to the central part of the histogram, shown with the purpose to indicate the strong deviation of the tails from the normal PDF. The PDF values between the two Gaussian curves are expected to correspond to magnetic structures, possibly flux tubes \citep{Greco09}.

\section{Summary and conclusions}
The main aim of this study was to show that the thin current sheets identified by simple structure detection tools populate the fat tails of PDFs.
This has already been shown for the histograms of the normalized current density in MHD turbulence simulations \citep{Greco09, Matthaeus15}.
However, in simulations the identification of current and magnetic structures forming the non-Gaussian PDF($\textbf{\textit{j}}$) is straightforward.
Although the four-point curlometer technique is an excellent tool for estimating $\textbf{\textit{j}}(t)$ from four-spacecraft data, it works well for selected short events only.
The two-point magnetic field differences have already been used previously in turbulence studies to obtain the long tail PDFs \citep{Voros06} and to identify the intermittency effects \citep{Yordanova15}.
The extremal values of magnetic field partial derivatives calculated between spacecraft pairs are also associated with thin current sheets.
We found that the strongest current sheets associated with $\partial_{ij}(extr)$ belong to the tails of PDF($\partial_{ij}(extr)$), confirming the results of MHD turbulence simulations on the generation of intermittent structures \citep{Greco09, Matthaeus15}. It was also found that during a one-hour-long time Cluster observed about a hundred thin magnetic structures in the MS downstream of a quasi-parallel BS which might be associated with non-homogenous localized heating of plasma. This conjecture has to be confirmed by a thorough analysis of similar events using high resolution field, particle and plasma data from the MMS mission. Our results also demonstrate that, a conditional selection of structures and their identification in histograms (PDFs) represents a powerful tool for better understanding of the role of rare but intense events (in our case strong discontinuities or reconnection sites), which can determine the basic physical properties of plasma systems.






\acknowledgments

Z.V. was supported by the Austrian Fond zur Förderung der wissenschaftlichen Forschung (project P24740-N27).
The research leading to these results has received funding from the European Community's Seventh Framework Programme ([$7/2007-2013$])
under grant agreement $n^\circ  313038$/STORM.

\clearpage

\begin{figure}
\centering
\includegraphics[width=20pc]{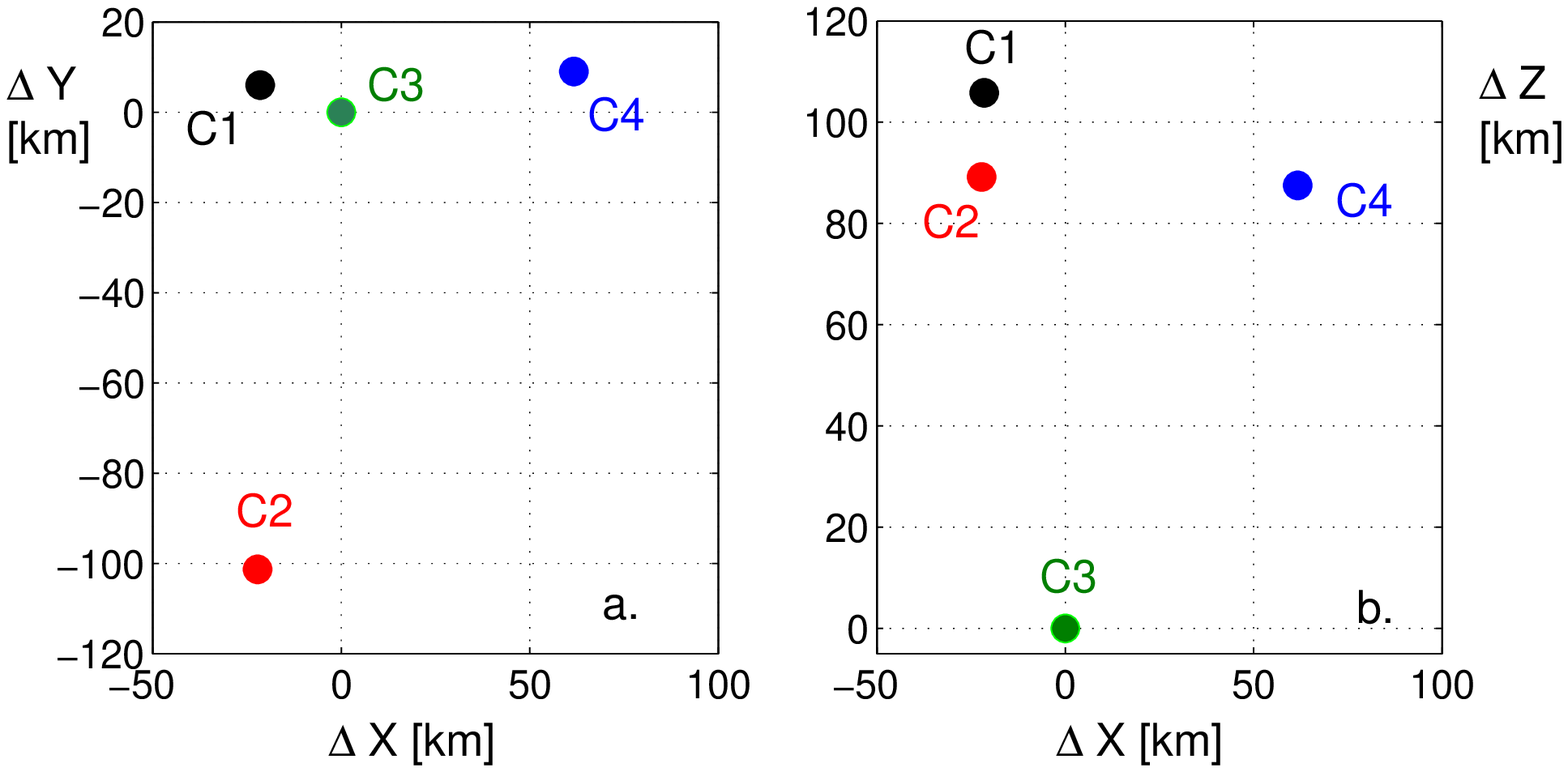}
\vspace{6cm}
\caption{Relative positions of the Cluster spacecraft. $a.$ $\Delta X$ vs. $\Delta Y$; $b.$ $\Delta X$ vs. $\Delta Z$.}
\label{fig1}
\end{figure}

\clearpage

\begin{figure}
\centering
\includegraphics[width=20pc]{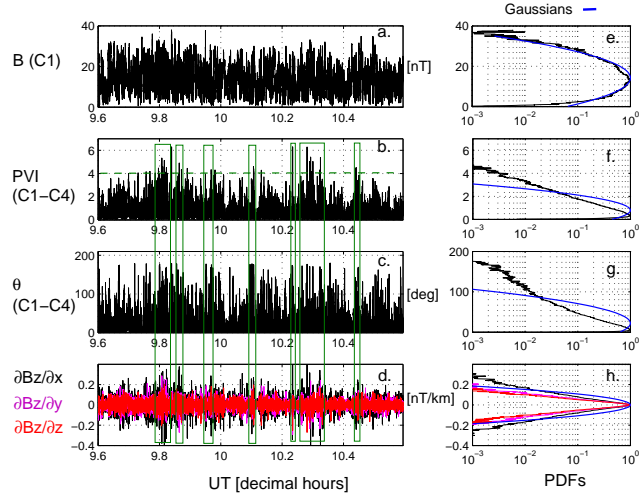}
\caption{Left panels, $a.$--$d.$: Magnetic field intensity for C1; PVI magnitude for C1--C4 with threshold $4$ (dashed green line); magnetic field rotation angle for C1--C4; and partial derivatives of magnetic field $B_Z$ component. The boxes outlined in green show examples of detected discontinuities above the given threshold PVI=4. Right panels, $e.$--$h.$: PDFs (histograms) corresponding to the data on the left. Gaussian distributions are shown in blue color.}
\label{fig2}
\end{figure}

\clearpage

\begin{figure}
\centering
\includegraphics[width=20pc]{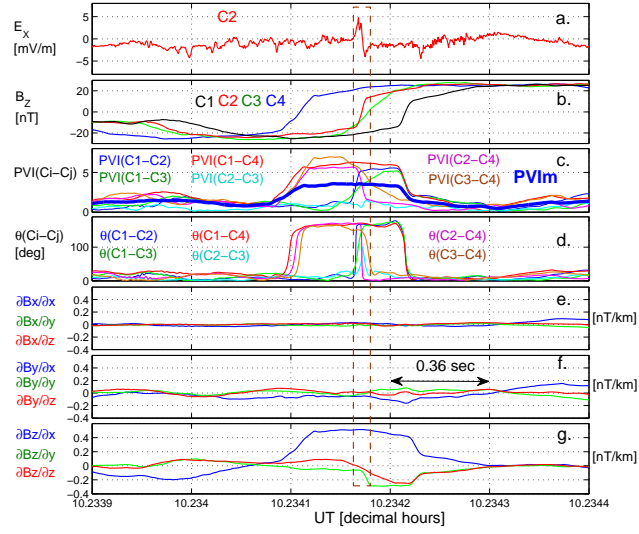}
\caption{An example of a current sheet. This event and the current associated electron heating has already been studied by \citet{Chasapis15a}. $a.$ Normal electric field $E_X$; $b.$ $B_Z$ components of the magnetic field; $c.$ Partial Variances of Increments: $PVI(Ci-Cj,t) \equiv PVI_{ij}(t)$ and $PVIm(t)$; $d.$ Magnetic field rotational angles $\theta(Ci-Cj,t) \equiv \theta_{ij}(t)$; $e.$--$g.$ Partial derivatives of magnetic field components.}
\label{fig3}
\end{figure}

\clearpage

\begin{figure}
\centering
\includegraphics[width=20pc]{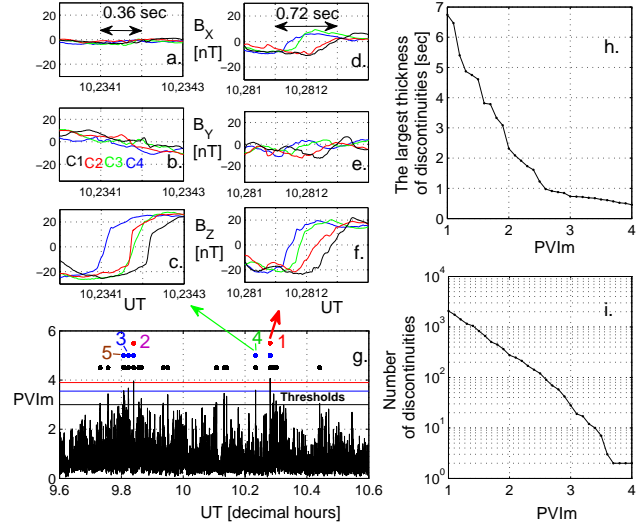}
\vspace{3cm}
\caption{Examples of current sheets with width $0.36$ s (panels $a.$--$c.$) and $0.72$ s (panels $d.$--$f.$); $g.$ PVIm magnitude with three levels of threshold (horizontal lines in red, blue and black) and their respective current sheets (colored dots). The five strongest discontinuities are color coded in red, magenta, blue, green and brown. The arrows mark the positions of the two selected events(1 and 4) in $a.$--$c.$ (previously studied by \citet{Chasapis15a} and $d.$--$f.$ (previously studied by \citet{Retino07}; $h.$ and $i.$ show the discontinuity thickness and number vs PVIm, respectively.}
\label{fig4}
\end{figure}

\clearpage

\begin{figure}
\centering
\includegraphics[width=20pc]{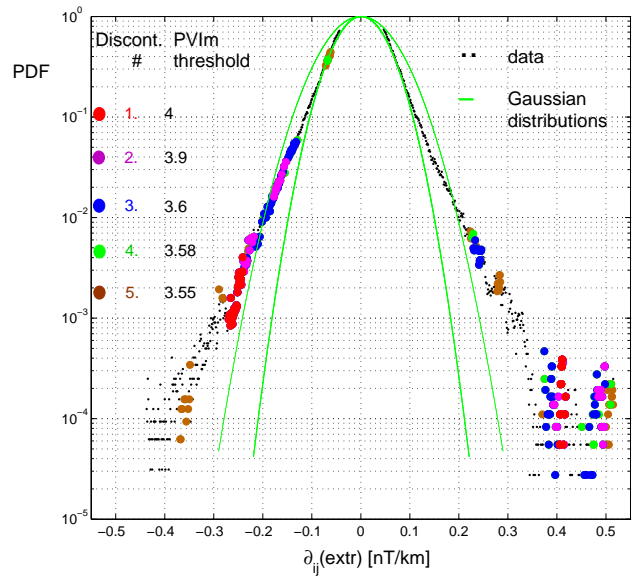}
\caption{PDF calculated from extremal values of magnetic field partial derivatives. The first five strongest discontinuities from Figure $4g.$ are color-coded points at the tail of PDF. Black points show the data, green lines correspond to Gaussian distributions.}
\label{fig5}
\end{figure}

\end{document}